\documentclass[copyright]{eptcs} 
\providecommand{\event}{To appear in PostML'2014} 
\usepackage{breakurl}             

\usepackage[T1]{fontenc}
\usepackage[latin1]{inputenc}
\usepackage{amsmath}
\usepackage{amssymb}
\usepackage{stmaryrd} 
\usepackage{array}
\usepackage{url}
\usepackage{verbatim}
\usepackage{xspace}
\usepackage{color}
\usepackage{yhmath}
\usepackage{mathrsfs}
\usepackage{mathtools}
\usepackage{textcomp}
\usepackage{doi}

\usepackage{fancyvrb} 
\def\mytilde{\kern -.15em\lower .6ex\hbox{\~{}}\kern .04em}

 \DefineVerbatimEnvironment
   {cod}{Verbatim}
   {commandchars=\{\}}

\definecolor{Cquestioncolor}{rgb}{.8,.0,.0}

\definecolor {colorgrey}{rgb}{.75,.75,.75} 

\newcommand{\E}[1] {{\em #1}}       

\newenvironment{ul}
{ \begin{list}
        {$-$}
        {
         \addtolength{\leftmargin}{-3pt}
         \addtolength{\topsep}{- \parskip}
         \setlength{\itemsep}{2pt}
    }}
{ \end{list} \vspace*{5pt} }

\newcommand{\cref}[1]{Chapter~\ref{#1}}

\newcommand{\Hdatasep} {\leadsto}

\newcommand {\Jmgshide}[1] {}

\usepackage{ntheorem}

\newcommand{\Hstar} {*}  

\newcommand{\Himpl} {\vartriangleright}

\newcommand{\Iassert}[1]{ }

\newcommand{\etal}{~\E{et al.}}
\usepackage{balance}
\usepackage[10pt]{moresize} 

\usepackage{color}
\definecolor{dktest}{rgb}{0.9,0.9,0.9}
\definecolor{ltblue}{rgb}{0,0.4,0.4}
\definecolor{dkblue}{rgb}{0,0.1,0.6}
\definecolor{dkgreen}{rgb}{0,0.2,0}
\definecolor{dkviolet}{rgb}{0.3,0,0.5}
\definecolor{dkred}{rgb}{0.5,0,0}
\definecolor{light-gray}{gray}{0.95}

\usepackage{listings}
\usepackage{lstcaml}

\def\hlinetext#1{\hbox to \hsize{\hrulefill\;\lower.3em\hbox{{#1}}\;\hrulefill}}
\def\hlinetexttiny#1{\hbox to \hsize{\hrulefill\;\lower.1em\hbox{{#1}}\;\hrulefill}}
\def\hlinenotext#1{\hbox to \hsize{\hrulefill\hrulefill}}

\newcounter{myexamples}

\newcommand{\insertexample}[1] {
\refstepcounter{myexamples}
\label{#1} 
\hlinenotext{}
\vspace{0.4em}
\noindent
  \begin{minipage}[c]{0.97 \textwidth}
  \lstinputlisting[language=MyCaml]{examples/#1_src.txt}
  \end{minipage} 

\vspace{-0.2em}
\noindent
 \begin{minipage}[c]{0.97 \textwidth}
  \noindent{\tiny Old error:}
  \lstinputlisting[language=MyCaml]{examples/#1_old.txt}
  \end{minipage}

\vspace{-0.2em}
\noindent
 \begin{minipage}[c]{0.97 \textwidth}
  \noindent{\tiny New error:}
 \lstinputlisting[language=MyCaml]{examples/#1_new.txt}
  \end{minipage}

\hlinenotext{} 
\vspace{0.5em}
}

\title{Improving Type Error Messages in OCaml}
\author{Arthur Chargu\'eraud
  \institute{Inria, Universit\'e Paris-Saclay, France}
  \institute{LRI, CNRS \& Univ. Paris-Sud, Universit\'e Paris-Saclay, France}
  \email{arthur.chargueraud@inria.fr}
}
\def\titlerunning{Improving Type Error Messages in OCaml}
\def\authorrunning{Arthur Chargu\'eraud}

\begin{document}

\maketitle


\begin{abstract}
Cryptic type error messages are a major obstacle to learning OCaml
or other ML-based languages.
In many cases, error messages cannot be interpreted without 
a sufficiently-precise model of the type inference algorithm.
The problem of improving type error messages in ML
has received quite a bit of attention over the past two 
decades, and many different strategies have been considered.
The challenge is not only to produce error messages that are 
both sufficiently concise and systematically useful to the programmer,
but also to handle a full-blown programming language
and to cope with large-sized programs efficiently.

In this work, we present a modification to the traditional ML type inference 
algorithm implemented in OCaml that, by significantly reducing the left-to-right bias, 
allows us to report error messages that are more helpful to the programmer.
Our algorithm remains fully predictable and continues to produce
fairly concise error messages that always help making some progress towards fixing 
the code. We implemented our approach as a patch to the OCaml compiler
in just a few hundred lines of code.
We believe that this patch should benefit not just to beginners, but also to 
experienced programs developing large-scale OCaml programs.
\end{abstract}


\section{Introduction}

Typing errors in OCaml are a major obstacle to the adoption of
the language, leading many potential adopters of OCaml to run away before 
they really get a chance to appreciate the language.
Yet, producing ``good'' error messages for realistic ML programs turns out 
to be an unexpectedly hard problem.

On the one hand, it is tempting to report everything that contributes 
to a type error. This strategy, however, can quickly lead to very 
verbose messages, so long that the programmer will not read them.
Interactive type-debugging sessions have been proposed as a solution to 
verbosity (e.g. \cite{chitil-01}), 
however they can be very time consuming for the programmer.
Another proposed approach is the extraction of minimal error 
slices~\cite{haack-wells-04}. Although quite appealing, 
error slices do not fully address the problem: it remains to select
which of those minimal slices to first present to the user,
and to select on which locations to attract the user's attention
when the slice involves locations spanning over remote lines of code.
Heuristics can be used for these selections
(e.g. \cite{zhang-myers-14}), however they make 
error messages less predictable (in the sense that an expert would
not be able to guess what error message the compiler would report
given an incorrect program), and they may attract the programmer's
attention away from the actual location of the error.

On the other hand, in order to produce a short error message, one 
cannot report all locations that contributes to a type error. Thus,
the type-checker must somehow make a decision on which information to select. 
In traditional type-checking algorithm, this selection is implicit
and results from the order in which unifications are performed. 
However, this order induces a problematic left-to-right bias,
well identified by the early research on type 
errors~\cite{bernstein-stark-95,yang-99,mcadam-98}.
For example, when the ``then'' branch and the ``else'' branch of a
conditional have incompatible types, a traditional type-checker will
systematically locate the error on the ``else'' branch, even though
the error might just as well originate in the ``then'' branch.
Worse, some error messages can only be understood with knowledge of the
internals of the type inference algorithm.

More generally, regardless of the order in which unifications are
represented and performed, the problem is to select what information
to report when detecting a path that corresponds to a unification 
conflict. Yet, without knowledge of the programmer's intention, 
the type-checker alone usually cannot decide which part of the
path is to blame. Here again, if the type-checker makes an arbitrary decision, 
it may attract the programmer's attention away from the place where 
the error is actually located, making the programmer waste precious time.




The motivation for this paper is to devise a system for reporting
type errors in ML programs that (1) produces messages that are always concise;
(2) eliminates as much as possible of the left-to-right bias associated with
traditional type-checking algorithms; (3) does so in an efficient
and predictable manner; (4) integrates smoothly into an existing
type-checker for a full-blown language.
In this paper, we show that, with a small number of carefully-crafted changes 
to the order in which unifications are performed by the type
inference engine of OCaml, and with appropriate processing of the 
conflicts that may arise from these unifications, we are able 
to generate messages that, we argue, 
provide the programmer with much more useful information for locating 
and fixing type errors, and also requires less knowledge of the type inference
algorithm. 

The main features of our approach are as follows:
\begin{ul}

\item Improved error messages for function applications; in particular,
better treatment of errors involving higher-order functions (e.g., \ca{List.fold})
and arithmetic operators (e.g., `\ca{+}' in place of `\ca{+.}'), as described in Section~\ref{sec:applications}.

\item Improved error messages for incompatible branches in conditional and
pattern matching constructs, as described in Section~\ref{sec:branches}.

\item Improved error messages for subexpressions that do not have the type
expected by the language construction in which they appear, e.g., a while loop 
condition that does not have type \ca{bool}, as described in Section~\ref{sec:subexpressions}.

\item Improved error messages for very common ML-specific errors:
missing `\ca{()}' arguments, missing `\ca{!}' operators,
or missing `\ca{rec}' keywords, as described in Section~\ref{sec:mlerrors}.

\end{ul}

Our approach is implemented as a patch to the OCaml compiler,
consisting of a few hundred lines of code.
The patched compiler offers a new flag, called \ca{-easy-type-errors}.
When this flag is activated, the compiler first runs the original
OCaml type-checking algorithm. If there is no error, then our modified
type-checking algorithm is not executed at all. Otherwise,
our patched compiler considers the first top-level definition
that fails to type-check and attempts to type-check it again,
this time using our modified algorithm, in order to produce 
the error message that will be reported to the user.
(Note that the top-level definition considered for re-type-checking 
could be one item from a local module definition.)

Thanks to our strategy of first running the original type-checking algorithm,
we are able to guarantee full backward compatibility:
any program that compiles with the traditional OCaml compiler also compiles 
with our patched compiler in \ca{-easy-type-errors} mode,
and moreover compiles without any runtime overhead.
Besides, since our modified type-checking algorithm is only slightly slower
than the original algorithm, we are able to report errors efficiently even
in large-scale programs.
For example, we tested the ability of our patched compiler to produce
error messages for bugs artificially introduced in
several-hundred lines long functions located in the middle 
of the implementation of the OCaml type-checker itself.

Our patch is compatible with nearly all of the OCaml language,
with GADTs and overloading of record fields as notable exceptions. 
Programs including GADTs 
can be compiled with the flag \ca{-easy-type-errors} activated,
however if a type error occurs in a top-level definition using
GADT features then the error message produced will not be helpful.
One exception is for format functions (e.g. ``printf''),
whose types now make use of GADTs in OCaml. We recognize
calls to such format functions and locally process 
them using the original type-checking algorithm.
Regarding overloading record field names, our patch does not support them
because the overloading resolution is sensitive to the order in which
unifications are performed, and this order is modified by our alternative
type-checking algorithm. (More generally, our patch expects the source code
to correctly type-check with the \ca{-principal} option of OCaml.)

Our patched compiler can be obtained and executed through the following commands.
%
\small
\begin{verbatim}
  git clone -b improved-errors https://github.com/charguer/ocaml.git
  cd ocaml
  ./configure && make world.opt
  ./ocamlc.opt -I stdlib -easy-type-errors foo.ml
\end{verbatim}
\normalsize

\section{Error reporting for ill-typed applications}
\label{sec:applications}

\paragraph{Towards a new algorithm}

To type-check a curried application, the OCaml type-checker first infers the
type of the function, extracts from it the types expected for the arguments,
and then unifies these types one by one with the types of the arguments provided
to the function.
This process introduces a significant left-to-right bias in the case
of polymorphic functions. Indeed, the unification of the first arguments 
may result in instantiation of polymorphic variables that may
later raise conflicts when processing the subsequent arguments.

Our new algorithm for type-checking applications instead works as follows.
First, we compute the most-general type of the function and 
of each of the arguments provided, independently. 
Then, we try to unify the types for the application, using the same code
as the original OCaml type-checker. However, if this unification process
raises an error, then we catch this error and rely on a new piece of code
for generating the error message. The message we produce systematically
locates the error on the entire application. It then reports
a table with a first column showing the types expected for the
arguments, and a second column showing the types of the arguments actually
provided.
Note that, in order to properly report the type of the arguments
when they include polymorphic components, we need to save copies
of the types before we start to unify them. We construct these
copies by computing the types or type schema associated with the arguments, 
following a process similar to the procedure involved for 
generalizing the type of the body of a let-binding.


The following example illustrates a case where the function passed to
\ca{List.map} operates on integers although a list of float values is provided.
The source code is followed by the error message produced
by the original OCaml compiler, then by the error message
produced by our patched compiler.

\insertexample{example_map_bad}

In the example above, the OCaml type-checker locates the error inside the 
second argument, more precisely on the constant \ca{2.0}.
An alternative algorithm that unifies the second
argument before the first one would report an error on the application
of the \ca{+} operator. Both choices are arbitrary: the type-checker
has no way of guessing the intention of the programmer.
In contrast, we choose to report the types expected by the
function face to face with the types provided to the function.
Note that, in particular, we report the type ``\ca{int -> int}'' 
inferred for the anonymous function; this information can be
quite useful for understanding the cause of the error.


\paragraph{Additional examples}

The table presentation of types involved in ill-typed applications
makes it easy to spot swapped arguments.
Consider for example the following code, 
where the two arguments provided to \ca{String.index} have been swapped.

\insertexample{example_index_swap}


Our new algorithm also makes it easy to spot arguments that are
swapped in higher-order functions. The following example illustrates
a case where the two arguments of the function passed to \ca{List.fold_left}
have been swapped. The 
OCaml type-checker produces a cryptic error message:
``The type variable 'a occurs inside 'a list''.
By constrast, our algorithm produces a table whose first row makes
it relatively easily to spot that the two arguments of the anonymous function
have been swapped.

\insertexample{example_fold_left_swap_app_2}

Our new type-checking algorithm for applications also greatly benefits
binary mathematical operators. Consider the classic example
of ``\ca{+}'' being used in place of ``\ca{+.}'' for adding
two float values. The OCaml type-checker
reports an error on the left operand, that is, an error located {\em before} 
the actual error, making the error very hard to debug for beginners.
In contrast, our error message makes it fairly obvious that the
operator used does not operate on the right types.

\insertexample{example_add_bad}

Similarly, our algorithm helps debugging the case of negative numbers
not surrounded by parentheses. Whereas the OCaml type-checker
only reports a complex error involving an arrow type, our message
makes it clear that the minus sign is interpreted as a binary operator
of two arguments, instead of being treated as a unary operator.

\insertexample{example_f_minus_one}

\paragraph{Applications with too many arguments}

When the number of arguments provided to a function exceeds the number of
expected arguments, we produce an error message including a specific sentence
telling that the function is being applied to too many arguments.
A simple example follows.

\insertexample{example_apply_too_many}

Producing such messages, however, leads to a complication in the case polymorphic
functions. Such functions may, when applied, produce a function, which may in 
turn be applied to additional arguments. A typical example is \ca{fst p x},
where \ca{p} is a pair of two functions and \ca{x} is an argument to be 
provided to the first of these two functions. Here, the function \ca{fst}
is applied to two arguments, even though the type of this function 
has a single arrow visible in its type (\ca{'a * 'b -> 'a}).
Although this programming pattern typically never occurs in code written
by beginners, we nevertheless try to produce useful error messages.

When the return type of a function consists solely 
of a type variable,
we cannot be certain that the function is applied to too many arguments.
Therefore, in such situations, we simply report in the message that the application is ill-typed,
and we also show the return type of the function, so that the programmer can investigate
why unification fails. The following example illustrates this situation.

\insertexample{example_apply_too_many_1}
%

Note that, the programmer can generally obtain much clearer error messages 
by naming the intermediate functions being produced. For example, 
if we replace in the program above the second line with 
``\ca{let f = fst p in let _ = f 2.0}'', then we obtain a clear
error message explaining that the function 
\ca{f} expects an \ca{int} but was applied to a \ca{float}.

In the future, we could try to also report how the return type of the function
has been instantiated. For the example above, we would report that \ca{'a} is being instantiated as \ca{int -> int}.

\paragraph{Extension for labelled and optional arguments}

We have extended our algorithm to support labelled arguments
and optional arguments featured by the OCaml language.
Note that our current implementation does not place arguments with
corresponding labels face to face; we plan to fix this in future work.
An example follows.

\insertexample{example_apply_labels}

\section{Error reporting for incompatible branches}
\label{sec:branches}

\paragraph{Conditionals}

To type-check a conditional expression, the traditional OCaml compiler first type-checks
the \ca{then}branch, and then it type-checks the \ca{else} branch.
If the latter does not have the same type as the former, an error
gets raised somewhere inside the \ca{else} branch.
This process thereby introduces a significant left-to-right bias:
errors are systematically reported in the \ca{else} branch
---even though, roughly half of the time, the error actually originates 
in the \ca{then} branch.
We eliminate this bias by type-checking the two branches independently,
before trying to unify the two resulting types.
In case of unification failure, we report
the type of both branches, not presuming of the one to blame.
The following example illustrates this case.

\insertexample{example_incompatible_else}

Although our approach eliminates most of the left-to-right bias,
some of it remains, due to side-effects associated with 
unification that may be performed while type-checking the first branch.
The next example illustrates such a situation: a variable \ca{x} 
of unknown type is used as an integer in the first branch and then used
as a float in the second branch. The error that our patched compiler reports
is, like with the original compiler, located in the second branch, 
even though the first branch could be blamed just as much.

\insertexample{example_if_propagate}

The example above is quite tricky to handle. In order to produce a 
more informative error message, one would need a significantly more
involved infrastructure for performing substitutions independently
in the various branches. For more details, we refer to the discussion 
of McAdam's proposal~\cite{mcadam-98} described in the related work section.

We believe that the form of left-to-right bias that remains in our type-checking
algorithm is much less severe than that associated with the original
algorithm. Indeed, the remaining bias only concerns free type variables 
being unified in the branches, and not the entire content of the branches.
In addition, such free type variables are typically associated to
local variables, such as function arguments. When facing a
typing conflict involving a variable, it is straightforward to 
assign a type annotation to this variable and to re-type-check the
code so as to obtain a useful error location.

\paragraph{Pattern matching}

The pattern-matching construct generalizes the conditional construct.
As before, we are able to significantly reduce the left-to-right bias by
first type-checking each of the branches independently, and only then unifying
the types of the branches one by one. 
In case of failure when unifying the type of the $(n+1)$-th branch
with the unified type of the $n$ first branches, we report these two 
types. We mention that the $(n+1)$-th branch is the first one that
does not unify, but we do not suggest that this branch is necessarily the
one to blame.

The following example illustrates the case of a pattern matching
where the first branch returns the integer zero (although the intention 
was to return the float zero),
the second branch unifies with the type \ca{int}, and the third 
branch produces a float and therefore conflicts with the previous branches.

\insertexample{example_match_incompat_branches_3}

The next example consists of a more complex scenario of a recursive
function with pattern matching.
The function is inferred by OCaml to return integers, because its first
branch returns the integer zero (instead of the intended float zero), triggering
a type error in the second branch at the place where the function is used.
Interestingly, our algorithm is able to type-check the two branches independently.
In particular, the second branch type-checks successfully because we
unify the pattern matching expression with its expected type only 
after all the branches have been unified together.
The error reported simply points out a mismatch between the result types
of the two branches.

\insertexample{example_match_incompat_branches_2}

\section{Error reporting for incompatible subexpressions}
\label{sec:subexpressions}

If a language construct expects a subexpression to be of a given type 
(e.g., loop conditions should have type \ca{bool}) 
but that the subexpression provided by the programmer does not have this type, 
then we produce a specific error-message, e.g. ``This expression is the 
condition of a while loop, so it should have type \ca{bool}, but it has 
type \ca{foo}.'' We find that such specific error messages are much easier 
to parse and to interpret than generic unification error messages, 
especially for beginners.
The idea of syntax-specific rules for type error messages is not new.
Yet, somewhat surprisingly, it does not seem to have been applied 
in the implementation of popular ML compilers, although we learned that
Ishii and Asai~\cite{ishii-asai-14} recently implemented such 
syntax-specific error messages in an external type debugging tool for OCaml.

To every syntactic construct we can produce specific error messages
for each of its subterm. We next present two representative examples.
The first one illustrates the case of an ill-typed loop condition.

\insertexample{example_while_bad_condition}

The next example involves a conditional without an else branch.
This example is particularly interesting because it consists of a program
that is perfectly valid from a semantics point of view, yet
fails to type-check. We have seen equivalent programs written in practice
by beginners using OCaml as a first programming language, and we have seen
them being completely stuck---it appears to be very hard for a beginner
to fix a program that is inherently correct!

The function considered takes two arguments and returns an ordered list
storing these two values. The code includes a redundant condition,
{\em a priori} harmless. However, the OCaml type-checker produces an
surprisingly incomprehensible error message:
``The variant type unit has no constructor \ca{::}''.
Thankfully, with our dedicated support for subexpression of
language constructs, we are able to identify the source of the problem
to be the missing else branch.

\insertexample{example_missing_else}

\section{Error reporting for ML-specific errors}
\label{sec:mlerrors}

\paragraph{Message for missing ``\ca{()}''.}

A typical mistake made by OCaml beginners is to forget the 
``\ca{()}'' argument after basic functions such as \ca{read_int}.
When they do so, beginners face a type error 
message that describes a conflict involving an arrow type.
Yet, these beginners, when they write their very first programs, typically
have no clue what the arrow type is ---they often don't even
know what a function is. Thus, we believe that it is useful
to report specific messages for missing unit arguments.

To that end, we detect unification errors arising from conflicts
between a type of the form ``\ca{unit -> t}'' (for some type \ca{t})
and another type that does not unify with the former. 
In such situation, we add to the error message the sentence:
``{You probably forgot to provide ``\ca{()}'' as argument somewhere.''
The following example shows a program with a call to \ca{read_int}
that is missing its unit argument. 

\insertexample{example_missing_unit_readint}

The next example shows another instance of a missing unit argument, 
this time on a call to the function \ca{print_newline}.
When OCaml is used without the strict-sequence flag, it reports that 
some arguments are missing, whereas we are able to be more
specific, reporting that the argument ``\ca{()}'' is missing.
When OCaml is used with the strict-sequence flag,
it reports the message: ``This expression has type \ca{unit -> unit} 
but an expression was expected of type \ca{unit}.'' This message, 
which involves an arrow type, typically does not help beginners at all.
We improve the situation by suggesting that the expression is missing
a unit argument.

\insertexample{example_missing_unit_newline}

More generally, every time the missing unit argument error is detected
on an expression expected to be of unit type, we drop 
the word ``somewhere'' from the message as we can be sure that the location 
reported corresponds to the place where ``\ca{()}'' is missing.

\paragraph{Message for missing ``\ca{!}''.}

The ``\ca{!}'' symbol in OCaml is often forgotten, 
not only by beginners with previous experience in imperative programming,
but also by experienced OCaml programmers.
With just a few lines of code, we can add a dedicated suggestion
at the end of the error messages, similarly to what we did for the missing
unit argument.
To that end, we detect unification errors arising from conflicts between a type
of the form ``\ca{t ref}'' (for some type \ca{t}) and another type that does
not unify with the former.
In such situation, we add to the error message the sentence:
``You probably forgot a `\ca{!}' or a `\ca{ref}' somewhere.''.
The following example 
illustrates a function call of the form ``\ca{print_int r}'', 
where \ca{r} is a reference not preceded by a `\ca{!}' operator.

\insertexample{example_ref_missing_bang}

The next example 
shows that it is important to leave the word ``somewhere'' in the
message, because if the expression of type ``\ca{int ref}'' is the result of a 
function call, then the ``\ca{!}'' could be missing either 
on the last line of the definition of the function, or at the call site.
In general, the type-checker cannot guess the intention of the programmer.

\insertexample{example_missing_bang_delayed}

\paragraph{Message for missing ``\ca{rec}''.}

It is a common mistake to forget the \ca{rec} keyword in OCaml, in particular 
because OCaml, unlike most other programming languages,
does not have recursive bindings by default.
We therefore find it worth adding 20 lines of code in the type-checker 
for detecting such mistakes and printing a dedicated message.

When reaching a variable that is unbound in the current typing context,
the traditional algorithm stops with the error ``unbound variable''.
We add an additional step, to check whether the variable has been
mistakenly bound by a ``\ca{let}'' instead of being bound by a ``\ca{let rec}''.
If so, we display a specific message to suggest adding the \ca{rec} keyword.
The following example 
illustrates such a scenario.

\insertexample{example_let_missing_rec} 

To implement the detection of missing \ca{rec} keywords, we extend 
the typing context with shadow bindings.
Each time we enter the body of a \ca{let} definition, we add the name
of the corresponding variable as a shadow entry to the typing context, so that
we can later test for the existence of such a shadow entry. In practice,
a shadow variable is simply represented as the name of this variable 
prefixed with a character disallowed in the syntax of identifiers.

\section{Related work}

There is a large literature on the production of better typing errors for ML.
A comprehensive survey of pre-2006 work can be found in Section 3 of 
Heeren's PhD thesis~\cite{heeren-05}
and Section 10 of Wazny's PhD thesis \cite{wazny-06}.
Below, we focus on closely-related work and recent work.

\paragraph{Unification order}

Several researchers have investigated modifications to the unification algorithm
in an attempt to obtain more intuitive error messages. In particular,
the goal is to eliminate or at least tame the left-to-right bias associated 
with the way unifications are traditionally performed.
To avoid the left-to-right bias, Bernstein and Stark~\cite{bernstein-stark-95},
and Yang~\cite{yang-99} propose to type subterms bottom-up, 
returning a type for the term and all of its free variables,
and to then try to unify the types of the free variables.
McAdam~\cite{mcadam-98} describes a technique
that is able to eliminate all left-to-right bias without going as far as
typing fully-open terms. To that end, McAdam proposes to type-check subterms 
bottom-up, returning a type and a substitution for flexible type variables, 
and to then try to unify the substitutions involved. 

In our work, we also type-check subterms bottom-up, and the subterms are 
processed relatively independently, even though we do not manipulate 
and compare substitutions explicitly but instead stick to a standard
global union-find data structure.
In particular, our implementation reuses all the data structures 
and infrastructure that already exist in the OCaml compiler.
By processing the subterms more independently,
we are able to remove nearly all of the left-to-right bias, even though,
as we have explained, there remain a few cases where unification 
leads to side-effects visible across subterms.
We find our approach to be a good compromise between the aim to treat 
subterms as independently as possible and the need for conciseness, 
predictability, simplicity and efficiency.

The possibility of performing unifications in different orders is also discussed
in by Lee and Yi~\cite{lee-yi-00}, who show how several existing algorithms 
($\mathcal{W}$, SML/NJ's algorithm, OCaml's algorithm, $\mathcal{H}$, $\mathcal{M}$)
can be viewed as particular instantiations of a general algorithm. 
It is likely the case that our strategy for type-checking applications
could also be described as an instantiation of this general algorithm.
However, Lee and Yi do not discuss which instantiations might be best
suited for error reporting, and do not discuss the treatment 
of n-ary applications and pattern matching.

\paragraph{Heuristic-based approaches}

Helium~\cite{heeren-al-03,hage-heeren-07}, by Hage and Heeren, is a type-checker 
designed to produce better error messages for a subset of the Haskell language.
The Helium type-checker builds the complete unification graph, and then relies
on a number of heuristics for reporting a probable cause
of the error. For example, it includes heuristic for detecting
permuted arguments; it relies on the proportion of constants 
in an inconsistent part of the unification graph to determines
which constant is considered to be the error; and it considers
that a more recent definition is a more likely source of error.

Helium reports ill-typed applications in a similar way as we do,
although with a slightly different presentation.
More precisely, Helium prints the type of the function being applied, and then 
prints the type of the function as it should have been in order for the
application to the given arguments to be well-typed.
As our work shows, it is not needed to construct the complete
unification graph for gathering this information.

Regarding the typing of branching constructs, Helium relies 
on the expected type of the result. Indeed, this type is
often provided as an explicit type annotation in Haskell,
unlike in OCaml.
That said, Helium does not appear to go as far as us 
in terms of reporting the types of the branches
in the case where no annotation is available.

\paragraph{Error slicing}

Error slicing denotes the process of extracting from a unification graph 
a set of minimal paths which could explain a unification error. Each path is 
minimal in the sense that it only includes nodes that actually
contribute to the error. Such a path can be reported to the user
as a set of locations.
Dinesh and Tip~\cite{tip-dinesh-01} introduce type error slicing 
in the context of an explicitly-typed languages.
Haack and Wells~\cite{haack-wells-04} apply error slicing
in the context of a ML language with type infererence,
and identify the criteria of completeness and minimality 
for type error slices.

The Chameleon tool~\cite{wazny-06} applies error slicing to Haskell,
addressing the issue of overloading, and also
attempting to reduce the number of locations reported.
Becker~\cite{becker-09} adapts Haack and Wells's algorithm to OCaml,
also with a few OCaml-specific extensions.
Kustanto and Kameyama~\cite{kustanto-kameyama-10} propose
algorithmic improvements to 
Haack and Wells's algorithm.
Zhang and Myers~\cite{zhang-myers-14} analyse the error slices 
produced using Bayesian
principles in order to identify the explanations most likely
to be correct, and thereby report a single location to the
user.
The Skalpel tool~\cite{rahli-al-15}, by Rahli\etal, builds a constraint
and extracts an error slice; it covers all of the SML programming
language, and overcomes efficiency challenges in dealing with
realistic programs.


\paragraph{Interactive debugging}

The basic idea of interactive debugging is to replace subexpressions
with a dummy token with flexible type, in order to investigate whether
this subexpression is responsible for a type error.
Bernstein and Stark~\cite{bernstein-stark-95} describe
this idea of interactive debugging, although they
suggested the user performing the change by hand in the code.
Chitil~\cite{chitil-01} describes a type-checker that features 
debugging sessions, during which the user is asked a sequence of
questions of the form: ``Is the intended type of expression \E{foo}
an instance of the type \E{bla}?'', to which the programmer
answers by either ``yes'' or ``no''. At the end of such a session, the
tool reports the expression that is to blame for the error.
Bra\ss el~\cite{brassel-04} automates the interactive debugging 
process in the TypeHope tool,
so as to produce error messages and suggestions for fixes.


Seminal~\cite{lerner-al-06,lerner-al-07}, by Lerner\etal, is a tool that
applies interactive debugging to the core OCaml language, making
use of the standard OCaml type-checker as a black-box.
Once the type-checker has produced a list of probable locations for the error,
Seminal relies on a number of heuristics for ranking the error messages and 
reporting the presumably most-useful ones first.
We ran Seminal on our set of example programs.
We observed that Seminal produces very useful error messages for 
swapped arguments, for missing "!" and missing "()".
However, for a number of programs, the error messages were not
so helpful, and sometimes worse than the error produced by 
the original OCaml type-checker. Moreover, every time the tool reports
not just a single cause but several plausible causes for the error, 
the error message was fairly long, which we felt could discourage
programmers or at least decrease their productivity.

One challenge with the interactive debugging approach is that 
it is not always clear how to handle programs containing multiple 
type errors.
Also, interactive debugging does not address the issue of left-to-right bias.
Furthermore, because it iterates calls to a type-checker 
(possibly viewed as a black-box), interactive debugging
may suffer from algorithmic inefficiencies when trying to produce 
error messages for errors located in large pieces of code.

\section{Conclusion}

We have presented a practical approach to improving type error messages
in OCaml, in particular by decreasing the amount of left-to-right bias, 
while preserving an efficient and predictable algorithm that produces 
concise messages.
Our implementation covers all the commonly-used features of the OCaml language. 
As we have argued, the approach scales up to large-scale programs with 
virtually no overhead. 
While our work has been mainly motivated by OCaml,
we believe that our approach to reporting errors for applications and
branching constructs could be similarly applied in other functional 
programming languages. 


\paragraph{Acknowledgements}
We are very grateful to Damien Doligez and Gabriel Scherer for 
interesting discussions and extensive feedback on this work,
and we thank the editor and reviewers for their very useful
comments.

\nocite{*}
\bibliographystyle{eptcs}
\bibliography{biblio}

\end{document}